# AN EFFICIENT ACCELERATOR DESIGN METHODOLOGY FOR DEFORMABLE CONVOLUTIONAL NETWORKS

*Saehyun Ahn[1], Jung-Woo Chang[1,2], and Suk-Ju Kang[1]*

[1]Department of Electronic Engineering, Sogang University, Seoul, South Korea
[2]LG Uplus, Seoul, South Korea
{hh585, zwzang91, sjkang}@sogang.ac.kr

## ABSTRACT

Deformable convolutional networks have demonstrated outstanding performance in object recognition tasks with an effective feature extraction. Unlike standard convolution, the deformable convolution decides the receptive field size using dynamically generated offsets, which leads to an irregular memory access. Especially, the memory access pattern varies both spatially and temporally, making static optimization ineffective. Thus, a naive implementation would lead to an excessive memory footprint. In this paper, we present a novel approach to accelerate deformable convolution on FPGA. First, we propose a novel training method to reduce the size of the receptive field in the deformable convolutional layer without compromising accuracy. By optimizing the receptive field, we can compress the maximum size of the receptive field by 12.6 times. Second, we propose an efficient systolic architecture to maximize its efficiency. We then implement our design on FPGA to support the optimized dataflow. Experimental results show that our accelerator achieves up to 17.25 times speedup over the state-of-the-art accelerator.

*Index Terms—* Hardware accelerator, deformable convolution, system architecture, FPGA, deep learning

## 1. INTRODUCTION

With the rapid development of deep learning, various applications, including image classification and object recognition, have shown a significant performance [1], [2]. Deep convolutional neural networks (CNNs) have been widely adopted in object detection models to extract features and predict multiple objects [3]-[6]. The architecture with several convolutional layers located in front of the object detection models is called the backbone network. The well-known backbone networks are VGGNet [7], ResNet [8], and ResNeXt [9]. As the role of the backbone network has been important, there are a lot of efforts to enhance performance [10]. Recently, Dai et al [11] propose dynamic convolutional networks, deformable convolutional networks (DCNs), to resolve the fixed geometric operations. This network consists of deformable convolutional layers (DCLs), which add 2D offsets to the regular grid sampling locations in the standard convolution. An example of DCL is depicted in Fig. 1. Output feature maps are obtained from weights and irregularly sampled input data obtained from offsets. However, due to learned offsets, the size of the receptive field is dynamically changed in both spatially and temporally.

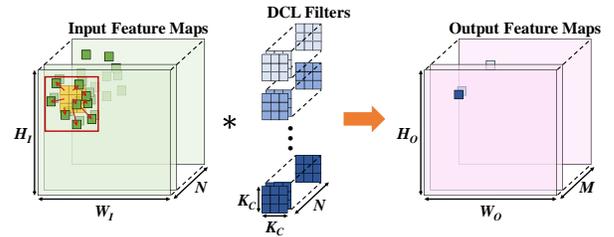

Fig. 1. Example of deformable convolution implementation with 3×3 kernel size. The red bounding box indicates the receptive field.

In an effort to deploy CNN models on hardware, various hardware accelerators have been proposed to improve the computation load for various types of layers [12]-[16]. Especially, these accelerators use FPGA as the target hardware because FPGA-based accelerators provide high energy efficiency and fast re-configurability [17]. However, when implementing the DCL with existing FPGA-based accelerators, DCL can cause some serious problems. First, DCL leads to irregular accesses to DRAM, which consumes more energy compared to the sequential DRAM accesses [18], [19]. Second, irregular DRAM accesses require more control logic to communicate the read traffic between DRAM and on-chip buffer [20]. Third, dynamically generated offsets can cause the pipeline stall if a cache miss occurs in input buffers. During this process, the processing elements (PEs) go into an idle mode, which results in a resource underutilization.

To address these challenges, we propose a novel training method that transforms the DCN into a hardware friendly model without compromising accuracy. Also, we present an accelerator architecture to implement the optimized model on FPGA. The main contributions of this paper are as follows.

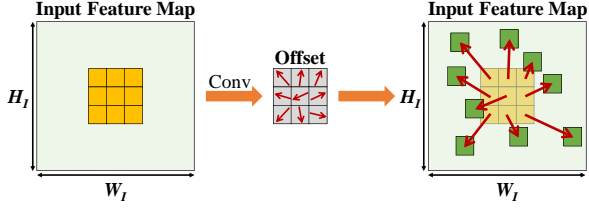

Fig. 2. Example of sampling process in the DCL with 3×3 kernel size. The red arrows indicate sampling directions and distances. The green boxes indicate sampling positions.

- A novel regularization algorithm has been proposed to transform a DCN into a hardware friendly model. This algorithm reduces the receptive field size significantly without performance degradation.
- An efficient architecture has been proposed to maximize the data reuse. This architecture is well optimized for a DCN by preventing the irregular DRAM accesses.

## 2. DEFORMABLE CONVOLUTIONAL LAYER ANALYSIS

DCL is a combination of two convolutional layers. In the first convolutional layer, offsets are generated to locate input data, which is formulated as follows,

$$\mathbf{o} = f(\mathbf{x}, \mathbf{w_o}) , \quad (1)$$

where $f(\cdot,\cdot)$ denotes a function for a convolution operation. $\mathbf{o}$, $\mathbf{x}$, and $\mathbf{w_o}$ denote tensors of offsets, input feature maps, and offset weights, respectively.

An example of the sampling process in the input feature map is depicted in Fig. 2. In the first layer, the positions of the inputs for the second layer are stored in offset tensors. Thus, the receptive field size of the second layer is determined by the maximum value of offset tensors, which causes low data reuse in hardware implementation. Then, the bilinear interpolation is performed to produce the inputs from real coordinates.

Then, the second convolution operation is derived as

$$\mathbf{y} = f(g(\mathbf{x}, \mathbf{o}), \mathbf{w_{deform}}) , \quad (2)$$

where $g(\cdot,\cdot)$ denotes a function for a sampling process, including bilinear interpolation at the sampling position. $\mathbf{y}$, and $\mathbf{w_{deform}}$ denote tensors of the output feature maps, and deformable convolutional weights, respectively.

## 3. DEFORMABLE CONVOLUTIONAL ACCELERATOR DESIGN

We present a novel method to reduce the receptive field size of the DCLs in section 3.1. In section 3.2, we propose an efficient accelerator of the DCLs trained by the proposed method in section 3.1.

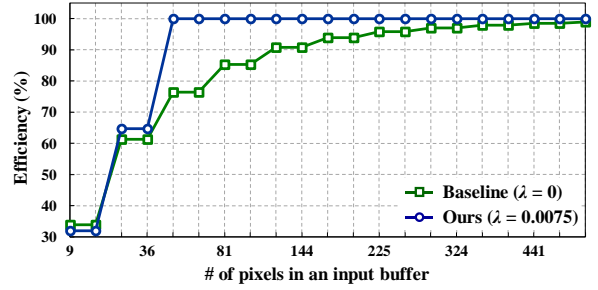

Fig. 3. The efficiency of the various input buffer capacity.

### 3.1. Receptive Field Optimization for Deformable Convolutional Layers

Let the maximum value of offsets are as follows,

$$o_{\max} = \max_{o_i \in \mathbf{o}} |o_i| , \quad (3)$$

where $o_{\max}$ is calculated with absolute values of offsets, because the sign means the sampling direction.

In the DCL, the size of the receptive field, $RF$, can be obtained by the following formulation.

$$RF = K_C + 2 \cdot \lceil o_{\max} \rceil , \quad (4)$$

where $K_C$ denotes the kernel size of DCL.

Fig. 3 shows an efficiency graph of the input buffer with various input buffer capacity. The efficiency is defined as the percentage of pixels that are read from the input buffer to compute the bilinear interpolation. If the pixels are not stored in the buffer, it causes the cache miss, which leads to high memory bandwidth. Thus, the input buffer needs at least 13.8MB when $\lambda$, which controls the receptive filed size of the DCL during training, is zero. However, large on-chip buffer systems cause high energy consumption. To prevent this issue, the receptive field size of DCL should be reduced.

The loss function using a novel regularization term is formulated as follows,

$$Loss = (1-\lambda) \cdot L + \lambda \cdot \max_{l \in \mathbf{D}} o^l_{\max} \text{ for } 0 \leq \lambda < 1, \quad (5)$$

where $L$ and $\mathbf{D}$ denote existing loss function and a set of DCLs in the network, respectively. The regularizer induces DCN to have a small receptive field size. Also, the regularizer is simple and incurs a negligible computational overhead, but powerful to transform a DCN into a hardware friendly model.

### 3.2. Architecture Design for Preventing Irregular Memory Accesses

Existing CNN architectures [21], [22] are inappropriate to accelerate DCL due to the random sampling process. The novel DCL engine prevents irregular memory accesses using the hardware friendly DCN. Additionally, the bandwidth between DRAM and on-chip buffers is dramatically decreased by increasing the number of data reuses.

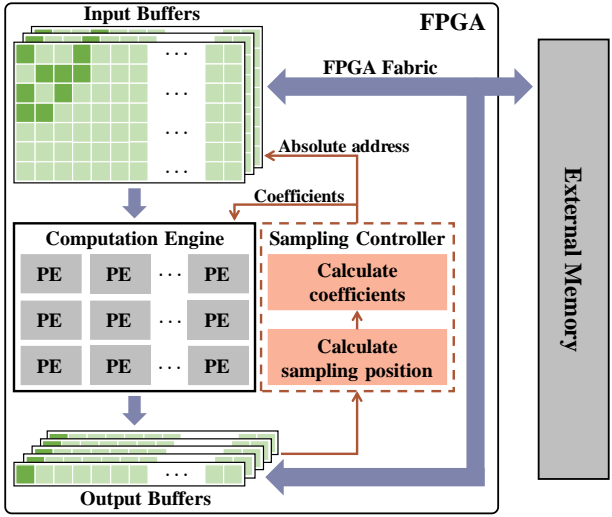

Fig. 4. The overall DCL accelerator.

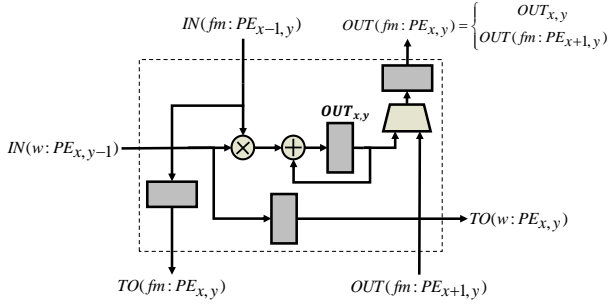

Fig. 5. Structure of $PE_{x,y}$ in systolic array architecture.

Fig. 4 shows an overall architecture of the proposed DCL accelerator. The DCL accelerator is mainly operated in two parts: input sampling stage and dynamic convolution stage.

In the input sampling stage, the computation engine, which convolves inputs with weights, creates the offset tensors and stores them in the output buffers. Then, the sampling controller calculates the sampling positions and coefficients. The computation engine executes bilinear interpolation with the coefficients after receiving data stored in the sampling position addresses from the input buffers. The interpolated inputs are saved in the output buffers and then transferred to DRAM. In addition, we reuse input blocks without DRAM access for offset tensors generation and bilinear interpolation. Therefore, excessive DRAM accesses are reduced.

After the input sampling stage, the dynamic convolution stage begins. The input buffer fetches the interpolated inputs and then uses them as the inputs of the second convolutional layer. The output tensors are generated in the computation engine and then saved in the output buffer. Finally, the output buffer transfers the output feature maps to DRAM. We design all the processes to be fully pipelined to achieve low latency.

In the process of computation and communication, we use a loop tiling method to leverage the hardware resources by dividing the feature maps into tiles [21]. $T_H$, $T_W$, and $T_N$ are height, width, and the number of channels of the input tiles, respectively. $T_M$ is the number of channels in the output tiles.

The input buffer size to prevent the irregular DRAM access overhead is defined as follows,

$$\text{Input buffer size} = RF \times (S \cdot T_W + RF - S) \times T_N, \quad (6)$$

where $S$ denotes stride. The output buffers are used to save the offsets, interpolated inputs, and output tensors. Therefore, the output buffer size is defined by the following formulation.

$$\text{Output buffer size} = T_W \times T_N \times 2 \times K_C^2. \quad (7)$$

PEs are designed based on the systolic array architecture [22], which efficiently reduces the bandwidth between buffers and PEs. Fig. 5 shows the $PE_{x,y}$, which is the PE architecture located at $(x, y)$ when the 2D systolic array.

To decide the tile sizes, we use a roofline model [21]. We set the attainable performance and computation to communication ratio for each stage. We can decide the optimal tiling factors using cross-layer optimization. Thus, we set $T_N$, $T_M$, $T_H$, and $T_W$ to 512, 64, 1, and 8, respectively.

TABLE I
PERFORMANCE EVALUATION ON THE VARIOUS $\lambda$ CONDITIONS

| $\lambda$ | $AP^b$ | $AP^b_{50}$ | $AP^b_{75}$ | $AP^b_S$ | $AP^b_M$ | $AP^b_L$ |
|---|---|---|---|---|---|---|
| 0 | 39.9 | 61.8 | 43.4 | 24.3 | 43.8 | 51.6 |
| 0.01 | 39.0 | 60.8 | 42.7 | 23.2 | 43.0 | 49.9 |
| 0.005 | 39.4 | 61.1 | 43.0 | 23.6 | 43.0 | 51.0 |
| **0.0075** | **39.3** | **61.1** | **42.7** | **23.5** | **42.8** | **50.6** |

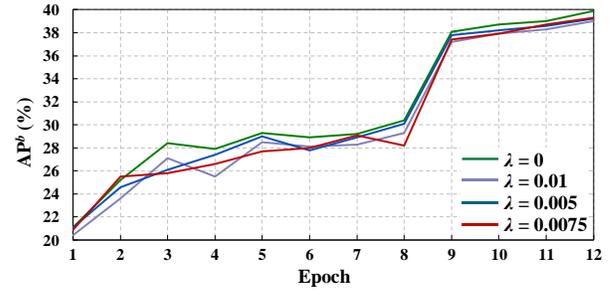

Fig. 6. Average precision comparison at various $\lambda$ conditions.

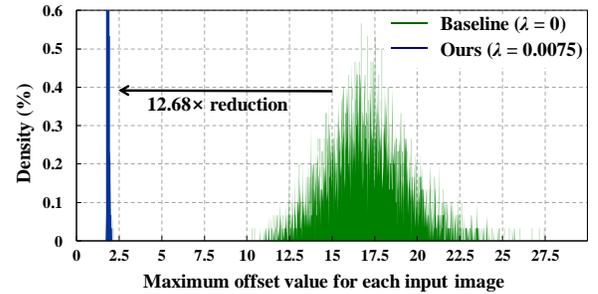

Fig. 7. Histogram of the maximum offset value in **D** on the various $\lambda$.

TABLE II
RESOURCE UTILIZATION FOR RESNET-50

|  | BRAM | DSP | LUT | FF |
|---|---|---|---|---|
| Ours | 601 | 2,596 | 298,249 | 356,612 |

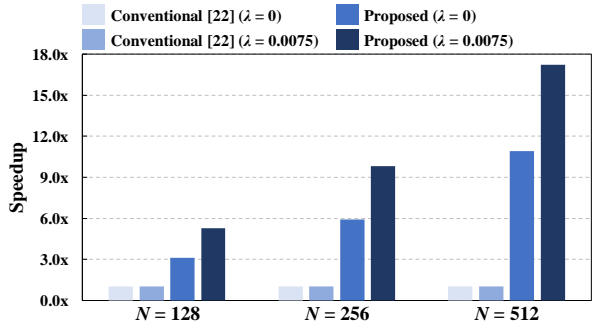

Fig. 8. Speed comparison of the proposed DCL accelerator and the state-of-the-art work [22]. The $N$ indicates the number of channels in the input feature maps.

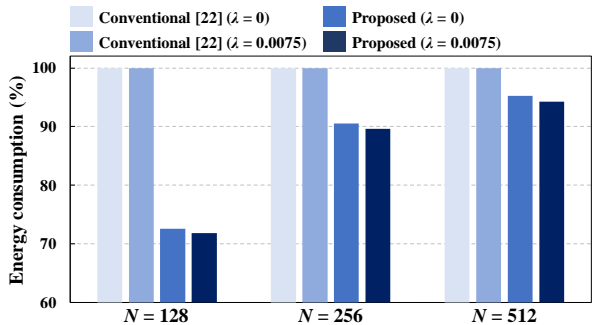

Fig. 9. Energy consumption comparison of the proposed DCL accelerator and the state-of-the-art work [22]. The $N$ indicates the number of channels in the input feature maps.

## 4. EXPERIMENTAL RESULTS

### 4.1. Evaluation of Receptive Field Optimization

We used Faster R-CNN [5] model, which uses 12 DCLs in ResNet-50 as a backbone network. We trained and evaluated using the MS-COCO 2017 dataset [23]. The validation dataset was only used for the test. In addition, our method was implemented based on the latest released version of MMDetection platform [24]. We used an NVIDIA Titan Xp GPU and trained for a total of 12 epochs with SGD optimizer, 0.005 learning rate, 0.9 momentum and 0.0001 weight decay.

Table I shows the object detection performance under the diverse $\lambda$ conditions. $AP^b$ represents box mean average precision. As a result, our regularizer achieved similar performance when compared with before. Fig. 7 shows the histogram of the maximum receptive field size in the set of DCLs for the validation dataset. Our regularizer reduced the receptive field size by 12.68 times. Hence, we could maximize the hardware efficiency with only 3% input buffer capacity.

### 4.2. Evaluation of Deformable Convolutional Layer Accelerator

We performed the simulations on the Xilinx Virtex7 485T FPGA using a single-precision floating-point. We converted the C/C++ code to HDL using Vivado HLS (v2016.4). Additionally, we conducted a pre-synthesis experiment with C/RTL co-simulation. DRAM was 1GB DDR3, and the bandwidth between on-chip buffers was 4GB/s. Table II shows the resource utilization of our design. DCL accelerator utilized a lot of LUTs and FFs in the sampling controller.

We derived the FPGA speed comparison of the proposed DCL accelerator and the conventional accelerator [22], which also uses a systolic array architecture, with various $\lambda$ conditions. We evaluated the performance of the DCLs in ResNet-50 by dividing three cases with $N$ size, as shown in Fig. 8. The conventional accelerator caused the irregular DRAM accesses, so the speed dropped significantly due to the pipeline stall. As $N$ increased, the performance of the DCL accelerator was improved by increasing the number of data reuses. The combination of our algorithm and accelerator was 17.25 times faster than the conventional accelerator by preventing the irregular DRAM accesses and pipeline stall.

Fig. 9 shows the energy consumption. When $\lambda$ was zero, the DCL accelerator achieved better energy efficiency than the conventional accelerator with a higher data reuse rate and lower irregular DRAM access overhead. Also, the combination of our algorithm and accelerator saved energy consumption 1.39 times over the conventional accelerator.

## 5. CONCLUSION

This paper proposes a novel design methodology to accelerate the DCL. First, we present a novel loss function with a new type of regularizer. Thus, we compress the maximum size of the receptive field by 12.6 times. Second, we propose an efficient DCL accelerator to prevent irregular DRAM accesses. We evaluate our design on Xilinx Virtex7 485T FPGA and achieve 5.28 times to 17.25 times higher throughput over the state-of-the-art accelerator.

## 6. ACKNOWLEDGMENTS

This research was supported by Samsung Electronics, a grant(19PQWO-B153369-01) from Smart road lighting platform development and empirical study on test-bed Program funded by Ministry of the Interior and Safety of Korean government and the MSIT(Ministry of Science and ICT), Korea, under the ITRC(Information Technology Research Center) support program(IITP-2020-2018-0-01421) supervised by the IITP(Institute of Information & communications Technology Planning & Evaluation).

# 7. REFERENCES

[1] A. Krizhevsky, I. Sutskever, and G. E. Hinton, "Imagenet classification with deep convolutional neural networks," In *NIPS*, pp. 1097-1105, 2012.

[2] A. Graves and J. Schmidhuber, "Framewise phoneme classification with bidirectional LSTM and other neural network architectures," In *IJCNN*, pp. 2047-2052, 2005.

[3] R. Girshick, J. Donahue, T. Darrell, and J. Malik, "Rich feature hierarchies for accurate object detection and semantic segmentation," In *CVPR*, pp. 580-587, 2014.

[4] R. Girshick, "Fast R-CNN," In *ICCV*, 2015.

[5] S. Ren, K. He, R. Girshick, and J. Sun, "Faster R-CNN: Towards real-time object detection with region proposal networks," In *NIPS*, pp. 91-99, 2015.

[6] K. He, X. Zhang, S. Ren, and J. Sun, "Mask R-CNN," In *ICCV*, pp. 2980-2988, 2017.

[7] K. Simonyan, and A. Zisserman, "Very deep convolutional networks for large-scale image recognition," *arXiv preprint arXiv:1409.1556*, 2014.

[8] K. He, X. Zhang, S. Hen, and J. Sun, "Deep residual learning for image recognition," In *CVPR*, pp.770-778, 2016.

[9] S. Xie, R. Girshick, P. Dollar, Z. Tu, and K. He, "Aggregated residual transformations for deep neural networks," In *CVPR*, pp.5987-5995, 2017.

[10] Z. Zou, Z. Shi, Y. Guo, and J. Ye, "Object Detection in 20 Years: A Survey," *arXiv preprint arXiv:1905.05055*, 2019.

[11] J. Dai, H. Qi, Y. Xiong, Y. Li, G. Zhang, H. Hu, and Y. Wei, "Deformable convolutional networks," In *ICCV*, pp.764-773, 2017.

[12] J.-W. Chang and S.-J. Kang, "Optimizing FPGA-based Convolutional Neural Networks Accelerator for Image Super-Resolution," In *ASP-DAC*, pp.343-348, 2018.

[13] J.-W. Chang, K.-W. Kang, and S.-J. Kang, "An Energy-Efficient FPGA-based Deconvolutional Neural Networks Accelerator for Single Image Super-Resolution," *IEEE Transactions on Circuits and Systems for Video Technology (TCSVT)*, vol.30, no. 1, pp.281-295, 2020.

[14] J.-W. Chang, K.-W. Kang, and S.-J. Kang, "SDCNN: An Efficient Sparse Deconvolutional Neural Network Accelerator on FPGA," *In DATE*, pp.968-971, 2019.

[15] J.-W. Chang, S. H. Ahn, K.-W. Kang and S.-J. Kang, "Towards Design Methodology of Efficient Fast Algorithms for Accelerating Generative Adversarial Networks on FPGAs", *In ASP-DAC*, pp.283-288, 2020.

[16] T. Geng, A. Lim T. Wang, C. Wu, Y. Li, A. Tumeo, S. Che, S. Reinhardt, and M. Herbordt, "UWB-GCN: Hardware Acceleration of Graph-Convolution-Network through Runtime Workload Rebalancing," *arXiv preprint arXiv:1908.10834*, 2019.

[17] J. Qiu, J. Wang, S. Yao, K. Guo, B. Li, E. Zhou, J. Yu, T. Tang, N. Xu, S. Song, Y. Wang, and H. Yang, "Going deeper with embedded fpga platform for convolutional neural network," In *FPGA*, pp. 26-35, 2016.

[18] TN-41-01: Calculating Memory System Power for DDR3, Micron, 2007. [Online]. Available: https://www.micron.com/-/media/Documents/Products/Technical%20Note/DRAM/TN41_01DDR3_Power.pdf

[19] B. Jacob, S. W. Ng, and D. T. Wang. "Memory systems: cache, DRAM, disk," Morgan Kaufmann Pub, 2007.

[20] T. Chen and G. E. Suh, "Efficient data supply for hardware accelerators with prefetching and access/execute decoupling," In *MICRO*, 2016.

[21] C. Zhang, P. Li, G. Sun, Y. Guan, B. Xiao, and J. Cong, "Optimizing FPGA-based accelerator design for deep convolutional neural networks," In *FPGA*, pp. 161-170, 2015.

[22] X. Wei, C. H. Yu, P. Zhang, Y. Chen, Y. Wang, H. Hu, Y. Liang, and J. Cong, "Automated systolic array architecture synthesis for high throughput CNN inference on FPGAs," In *DAC*, pp. 1-6, 2017.

[23] T.-Y. Lin, M. Maire, S. Belongie, J. Hays, P. Perona, D. Ramanan, P. Dollar, and C. L. Zitnick, "Microsoft COCO: Common objects in context," In *ECCV,* pp. 740-755, 2014.

[24] K. Chen, J. Wang, J. Pang, Y. Cao, Y. Xiong, S. S. Xiaoxiao Li, W. Feng, Z. Liu, J. Xu, Z. Zhang, C. Z. Dazhi Cheng, T. Cheng, Q. Zhao, B. Li, X. Lu, R. Zhu, J. D. Yue Wu, J. Wang, J. Shi, W. Ouyang, C. C. Loy, and D. Lin, "MMDetection: Open mmlab detection toolbox and benchmark," *arXiv preprint arXiv:1906.07155*, 2019.